\documentclass[doublecol,comment]{epl2} 

\usepackage{amsmath}
\usepackage{graphicx}
\usepackage{amsfonts}
\usepackage{amssymb}

\usepackage[english]{babel}

\usepackage{hyperref}

\usepackage[T1]{fontenc}
\usepackage{lmodern}
\usepackage{microtype}

\newcommand{\D}{\ensuremath{\mathrm{d}}}

\newcommand{\ie}{\textit{i.\,e.,\@}}

\newcommand{\cf}{\textit{cf.\@}}

\def\fig#1{\ref{fig:#1}}
\def\Fig#1{Fig.\@~\fig{#1}}

\def\eq#1{(\ref{eq:#1})}
\def\Eq#1{Eq.\@~\eq{#1}}

\title{Comment on ``A test-tube model for rainfall''\\by Wilkinson M., EPL 106 (2014) 40001}

\author{Martin Rohloff$^{1,2}$ \and Tobias Lapp$^{1}$ \and J\"urgen Vollmer$^{1,2}$}
\institute{
$^{1}$ Max-Planck-Institut f\"ur Dynamik und Selbstorganisation (MPIDS), 37077 G\"ottingen, Germany\\
$^{2}$ Fakult\"at f\"ur Physik, Universit\"at G\"ottingen, 37077
  G\"ottingen, Germany}


\begin{document}
\maketitle

In Ref.~\cite{Wilkinson2014} Michael Wilkinson revisits a model for
rainfall \cite{Wilkinson2011,LappPhD2011} that is based on a crossover
from diffusive, Ostwald-like, growth for small droplets to growth
dominated by gravitational collisions where large sedimenting
droplets grow by collecting smaller ones.  Wilkinson asserts that the model is fully compatible with
our data \cite{LappPhD2011,LappRohloffVollmerHof2012}. Here we point
out why we can not support this conclusion.

We start our discussion from equation (11) of \cite{Wilkinson2014},
\begin{equation}
 \frac{\D a}{\D t} 
 = \delta\, \frac{D\Lambda}{a^2} 
 + \frac{\varepsilon}{4} \, \kappa a^2 \xi t
\label{eq:time-evolution}
\end{equation}
The second summand contributing to the growth of $a$ 
accounts for growth by gravitational collisions, where
$\xi$ characterises the
steepness of the temperature ramp, $\kappa$ is the prefactor
entering the Stokes law of droplet sedimentation, and $\varepsilon$ is 
the collection efficiency. It is expected
to take values in the range $0 < \varepsilon \leq 1$ \cite{BeardOchs1993}. 
The other summand accounts for the growth of small droplets of radius
$a$ by mass diffusion with diffusion constant~$D$. In this term, the
Kelvin length~$\Lambda$ accounts for effects from the surface tension,
and we added here a factor~$\delta$ which arises as an estimate for large $a$
of the term $(a/ a_0) - 1$ in Eq.~(8) of Wilkinson's paper \cite{Wilkinson2014}.
The Lifshitz-Slyozov theory \cite{LifshitzSlyozov1961,Bray1994}
asserts that $0 \leq a / a_0 \leq 3/2$ such that $\delta \lesssim 1/2$
for the largest droplets in the system. 
% Based on an argument presented three lines above Eq.~(9) 
Wilkinson uses $\delta=1$ in his equation~(11)
for the time evolution of the radius $a$.
For any choice of $\delta$ that is of order one, as required by
compatibility with Ostwald ripening in the initial stages of droplet
growth, we will now demonstrate that predictions based on
\Eq{time-evolution} are in variance with our data.

\begin{figure}[tbp]
\centering
  \includegraphics[angle=-90]{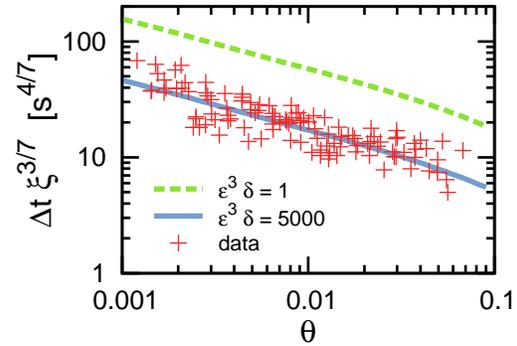}
  \caption{Data of the oscillation periods, $\Delta t$, of the
    isobutoxyethanol-rich phase of an isobutoxyethanol/water system
    \cite{LappPhD2011,LappRohloffVollmerHof2012}. The lines show the
    predictions of \Eq{DeltaT}. It only depends on the dimensionless
    parameter $\varepsilon^3 \, \delta$ -- the temperature dependent
    values of the other material parameters, $D$, $\Lambda$ and
    $\kappa$ entering \Eq{DeltaT} are known from independent
    measurements~\cite{LappPhD2011}. }
  \label{fig:test}
\end{figure}

The period of time $\Delta t$ needed for the largest droplets in the mixture to grow
from $a=0$ to $a=\infty$ is best obtained by non-dimensionalising
\Eq{time-evolution} with respect to the bottleneck time $t_1$ and the
bottleneck radius $a_1$ [\cf~Wilkinson's Eqs.~(12) and (13)],
\begin{equation} 
  t_1 = \left(\frac{4^3 / 3^4}{\delta D\Lambda (\varepsilon \kappa \, \xi)^3}\right)^{1/7} \, ,
\quad
 a_1 =\left(\frac{12 \delta^2 D^2 \Lambda^2}{\varepsilon \kappa \xi}\right)^{1/7}
\label{eq:bottleneck}
\end{equation} 
and numerical integration.
The resulting prediction 
\begin{equation}
 \Delta t 
 = 2.44 \; t_1 
 = 2.44 \left(\frac{4^3 / 3^4}{ \delta D\Lambda (\varepsilon \kappa \, \xi)^3}\right)^{1/7} \, ,
\label{eq:DeltaT}
\end{equation}
is tested in \Fig{test} where we plot $\Delta t \, \xi^{3/7}$ as
function of the reduced temperature,~$\Theta = |T - T_c|/T_c$.  We
note that the temperature dependence of the model parameters $D$,
$\Lambda$ and $\kappa$ has been provided in \cite{LappPhD2011}.
Therefore, $\varepsilon^3 \, \delta$ is the only free parameter of the
prediction, \Eq{DeltaT}.  The dashed green line shows the prediction
for the case $\varepsilon^3 \, \delta = 1$ which has been used in
\cite{Wilkinson2014}.  The physical bounds $0 < \varepsilon < 1$ and
$0 \leq \delta \lesssim 1/2$ derived above rather require
$\varepsilon^3 \, \delta \lesssim 1/2$. In that case the prediction is
shifted towards still larger values, \ie~further away from the
experimental observations.  A good fit of the data is obtained for
$\varepsilon^3 \, \delta \simeq 5000$ (solid blue line in \Fig{test})
that clearly lies out of the physical bounds.
We hence hold that \Eq{DeltaT} does not faithfully describe our data:

\textbf{1.} The dashed green line shows a lower bound to the best fit to the data when observing the
physical constraints $\varepsilon < 1$ and $\delta < 1/2$. Taking the
maximum value, $\delta < 1/2$, consistent with the Lifshitz-Slyozov
theory of Ostwald ripening requires collision efficiencies of the order of
$\varepsilon \simeq 20$ to arrive at a faithful description of the data
(solid blue line in \Fig{test} for $\varepsilon^3\delta \simeq 5000$). 
This value appears to be unrealistically large, in particular because we
do not expect turbulence enhancement of the collision
efficiencies~\cite{GrabowskiWang2013} for our experiments where
Re$\,\lesssim 10^{-1}$.

\textbf{2.} Our experimental data on the time evolution of the number
density $n$ suggest bottleneck radii of $a_1 \simeq 8\,\mu$m
(\cf~figure~3.18 in~\cite{LappPhD2011}). 
As expected this value is larger than the radius $a_S \simeq
1\,\mu$m where the the Stokes settling velocity overtakes the
displacement by diffusivity of the droplets, $D/a_S$.
On the other hand, for $\varepsilon^3 \, \delta = 5000$ and $\delta <
1/2$ equation~\eq{bottleneck} predicts radii $a_1$ of about $100\,$nm. 
% and one can easily check that 
For  $0 \leq \delta \lesssim 1$ one only obtains 
physically meaningful bottleneck radii for values of $\varepsilon$ where 
$\varepsilon \ll 10^{-6}$
% , unless one relaxes the constraint $0 \leq \delta
% \lesssim 1$.

\section{Conclusion}
When applied to our data the prediction, \Eq{DeltaT}, cannot cope with
the competing requirements of $\varepsilon \simeq 20$ and $\varepsilon \ll
10^{-6}$.  The dashed green line in \Fig{test} clearly shows that the
model suffers from the same \emph{quantitative} difficulties when
applied to the formation of terrestrial rainfall (as acknowledged in
\cite{Wilkinson2014}) and to the experimental data on binary demixing
that have been discussed in the present comment: For physically
realistic values of $\varepsilon$ and $\delta$, where $\varepsilon^3 \delta
\ll 1$, the prediction \Eq{DeltaT} provides values $\Delta t$ that are
substantially larger than those observed in experiments.
We hence believe that Ostwald ripening is not only too slow to account
for terrestrial rainfall, but it is also too slow to account for
rainfall in our experiments.

% \bibliographystyle{eplbib}
% \bibliography{comment}

\end{document}